%% file: brandt.tex
\documentclass[proceedings]{stacs} 
\stacsheading{2010}{107-118}{Nancy, France}
\firstpageno{107}

\usepackage{array,xspace,multirow,tikz}

\newcommand{\acknowledgements}{\section*{Acknowledgements}}

\usepackage[numbers]{natbib}
\usepackage{optparams,ifthen}
\makeatletter
\let\citeptemp\citep
\long\def\citep@[#1][#2]#3{%
	\ifthenelse{\equal{#1}{}}{%
		\ifthenelse{\equal{#2}{}}{%
			\citeptemp[][]{#3}}{%
			(\citeptemp[][]{#3}, #2)}}{%
		\ifthenelse{\equal{#2}{}}{%
			(#1\citeptemp[][]{#3})}{%
			(#1\citeptemp[][]{#3}, #2)}}}
\renewcommand{\citep}{\optparams{\citep@}{[][]}}
\makeatother

\usepackage[centertags]{amsmath}
\usepackage{amssymb,amsfonts}

\newcommand{\secref}[1]{Section~\ref{#1}}
\newcommand{\figref}[1]{Figure~\ref{#1}}

\newcommand{\thmref}[1]{Theorem~\ref{#1}}

\newcommand{\defref}[1]{Definition~\ref{#1}}

\DeclareMathSymbol{\Gamma}{\mathalpha}{letters}{"00}
\DeclareMathSymbol{\Delta}{\mathalpha}{letters}{"01}
\DeclareMathSymbol{\Theta}{\mathalpha}{letters}{"02}
\DeclareMathSymbol{\Lambda}{\mathalpha}{letters}{"03}
\DeclareMathSymbol{\Xi}{\mathalpha}{letters}{"04}
\DeclareMathSymbol{\Pi}{\mathalpha}{letters}{"05}
\DeclareMathSymbol{\Sigma}{\mathalpha}{letters}{"06}
\DeclareMathSymbol{\Upsilon}{\mathalpha}{letters}{"07}
\DeclareMathSymbol{\Phi}{\mathalpha}{letters}{"08}
\DeclareMathSymbol{\Psi}{\mathalpha}{letters}{"09}
\DeclareMathSymbol{\Omega}{\mathalpha}{letters}{"0A}
\DeclareMathSymbol{\thetaorg}{\mathalpha}{letters}{"12}

\newcommand{\ie}{i.e.,\xspace}
\newcommand{\eg}{e.g.,\xspace}
\newcommand{\cf}{cf.\@\xspace}

\newcommand{\parikh}{\#}
\newcommand{\numplayers}{n}
\newcommand{\numactions}{k}
\newcommand{\nat}{\mathbb{N}}

\newcommand{\midd}{:}

\newcommand{\me}{\textsf{ME}\xspace}
\newcommand{\mc}{\textsf{CE}\xspace}

\newcommand{\map}{\textsf{MP}\xspace}
\newcommand{\ide}{\textsf{IDE}\xspace}
\newcommand{\ids}{\textsf{IDS}\xspace}

\begin{document}

\title{On Iterated Dominance, Matrix Elimination, and Matched Paths}
\author[lmu]{F. Brandt}{Felix Brandt}
\address[lmu]{%
  Institut f{\"u}r Informatik,
  Ludwig-Maximilians-Universit{\"a}t M{\"u}nchen,
  80538 M{\"u}nchen, Germany}
\email{{brandtf,fischerf}@tcs.ifi.lmu.de}
\author[lmu]{F. Fischer}{Felix Fischer}
\author[ug]{M. Holzer}{Markus Holzer}
\address[ug]{%
	Institut f{\"u}r Informatik, 
	Universit{\"a}t Gie\ss{}en,
	35392 Gie\ss{}en, Germany}
\email{holzer@informatik.uni-giessen.de}


\begin{abstract}
  We study computational problems arising from the iterated removal of weakly dominated actions in anonymous games.  Our main result shows that it is NP-complete to decide whether an anonymous game with three actions can be solved via iterated weak dominance.  The two-action case can be reformulated as a natural elimination problem on a matrix, the complexity of which turns out to be surprisingly difficult to characterize and ultimately remains open.  We however establish connections to a matching problem along paths in a directed graph, which is computationally hard in general but can also be used to identify tractable cases of matrix elimination.  We finally identify different classes of anonymous games where iterated dominance is in P and NP-complete, respectively.
\end{abstract}

\maketitle

\section{Introduction}

An \emph{anonymous game} is characterized by the fact that players do not distinguish between other players in the game, \ie their payoff only depends on the number of other players playing the different actions, but not on their identities.  Anonymous games constitute a very natural class of multi-player games which is also highly relevant in practice~\citep[\cf][]{DaPa08a}.  \emph{Symmetric} games additionally have identical payoff functions for all players.  A \emph{strategy} of a player is a probability distribution over his actions, and we say that an action is \emph{weakly dominated} if there exists a strategy of the same player guaranteeing him at least the same payoff for any combination of strategies of the other players, and strictly more payoff for some such combination.\footnote{Some authors~\citep[\eg][]{GKZ93a,KPT88a} use the terms weak dominance or dominance to refer to a weaker notion that does \emph{not} require the dominating strategy to sometimes yield a strictly higher payoff. This notion is called very weak dominance by other authors~\citep[\eg][]{Apt04a,BBFH09b}.}  
Dominated actions may be discarded for the simple reason that the player will never face a situation where he would benefit from using these actions.  The solution concept of \emph{iterated dominance} works by removing a dominated action and applying the same reasoning to the reduced game~\citep[\eg][]{Myer91a}.  A game is then called \emph{solvable} by iterated dominance if there is a sequence of eliminations that leaves only one action for each player.  Interestingly, anonymous games often arise in the context of voting, where dominance solvability was originally introduced~\citep{Moul79a}. 

Unlike iterated \emph{strict} dominance, which requires the dominating action in each step to be strictly better for \emph{every} combination of strategies of the other players, proper epistemic foundations for iterated \emph{weak} dominance are fairly hard to come by~\citep[\eg][]{BFK08a,Samu92a}.  Nevertheless, iterated weak dominance is an established and well-studied solution concept that occurs in virtually every textbook on game theory.  Its computational properties, however, are not well understood, particularly in restricted classes like anonymous games.  Potential computational hardness of iterated weak dominance stems from the fact that the result of the elimination process generally depends on the order in which actions are eliminated. 

\medskip\noindent\textbf{Related Work.}
Deciding whether a game in normal form can be solved by iterated weak dominance is NP-complete already for games with two players and two different payoffs and when restricted to dominance by pure strategies~\citep{GKZ93a,CoSa05b}.  In two-player constant-sum games, both solvability and eliminability of a given action become tractable, while reachability of a subgame remains NP-complete~\citep{BBFH09b}. The corresponding problems for \emph{strict} dominance can generally be solved in polynomial time~\citep{CoSa05b}. 

All of the above results concern games with few players and an unbounded number of actions. Unlike general normal-form games, anonymous and symmetric games allow for a succinct representation even when the number of players is unbounded.  Computational aspects of these games, particularly with respect to Nash equilibrium, have recently come under increased scrutiny due to their importance in modeling large anonymous environments like the Internet.  A Nash equilibrium of a symmetric game can be found in polynomial time if the number of actions is not too large compared to the number of players~\citep{PaRo05a}.  In the larger class of anonymous games, Nash equilibria admit a polynomial-time approximation scheme when there is only a constant number of actions~\citep{DaPa08a}.  The pure equilibrium problem is tractable in anonymous games with a constant number of actions, and NP-complete if the number of actions grows in the number of players~\citep{BFH09a}. 

\medskip\noindent\textbf{Results and Paper Structure.}
We begin by introducing the relevant game-theoretic concepts.  In \secref{sec:initial} we show that iterated dominance solvability is NP-hard for symmetric games with an unbounded number of actions, and tractable for symmetric games with a constant number of actions.  The rest of the paper is then concerned with the only remaining class, anonymous games with a constant number of actions.  In \secref{sec:matrix}, we show how the two-action case can be reformulated as a natural elimination problem on a matrix.  The complexity of this problem remains open, but in \secref{sec:mp} we draw connections to a matching problem on paths of a directed graph.  The latter problem, which may be of independent interest, is intractable in general but allows us to obtain efficient algorithms for restricted versions of matrix elimination.  In \secref{sec:three} we finally use the matching formulation to show NP-hardness of iterated dominance in anonymous games with three actions. 
Proofs are omitted due to space constraints, and will be given in the full version of the paper.

\section{Preliminaries}

An accepted way to model situations of strategic interaction is by means of a \emph{normal-form game}~\citep[\eg][]{Myer91a}.
\begin{definition}[normal-form game]
  A \emph{game in normal-form} is a tuple $\Gamma=(N,(A_i)_{i\in N},(p_i)_{i\in N})$, where~$N$ is a finite set of \emph{players} and for each player $i\in N$, $A_i$ is a finite set of \emph{actions} available to player $i$ and $p_i:(\prod_{i\in N}A_i)\rightarrow\mathbb{R}$ is a function mapping each action profile, \ie each combination of actions, to a real-valued \emph{payoff} for player~$i$.
\end{definition}
We write $S_i=\Delta(A_i)$ for the set of (mixed) \emph{strategies} of player $i\in N$, \ie the set of probability distributions over his actions, and call a strategy \emph{pure} if it selects some action with probability one.  A vector $s\in\prod_{i\in N}S_i$ will be called a \emph{strategy profile}.  Payoff functions naturally extend to strategy profiles, and we write $p_i(s)$ for the \emph{expected} payoff of player~$i$ in strategy profile~$s$.  We further write $\numplayers{}=|N|$ for the number of players in a game, $s_i$ for the $i$th element of strategy profile $s$, and $s_{-i}$ for the vector of all elements of~$s$ but $s_i$.

We will henceforth concentrate on games where $A_i=A$ for all $i\in N$ and some set~$A$.  Such a game is \emph{anonymous} if the payoff of player~$i$ is invariant under any automorphism~$\pi':A^N\rightarrow A^N$ of the set of actions profiles induced by a permutation~$\pi:N\rightarrow N$ of the set of players that satisfies $\pi(i)=i$~\citep[\eg][]{BFH09a}.
An intuitive way to describe anonymous games is in terms of equivalence classes of the automorphism group of~$\pi'$, using a notion introduced by \citet{Pari66a} in the context of context-free languages.  Given a set~$A$ of actions, the \emph{commutative image} of an action profile $a_N\in A^N$ is given by $\parikh(a_N)=(\parikh(a,a_N))_{a\in A}$ where $\parikh(a,a_N) = |\{\,i\in N\midd a_i=a\,\}|$. In other words, $\parikh(a,a_N)$ denotes the number of players playing action~$a$ in action profile~$a_N$, and $\parikh(a_N)$ is the vector of these numbers for all the different actions.  This definition naturally extends to action profiles for subsets of the players.  We consider four types of anonymity~\citep[\cf][]{BFH09a}. 
\begin{definition}[anonymity]\label{def:anonymity}
  Let $\Gamma=(N,(A_i)_{i\in N}, (p_i)_{i\in N})$ be a normal-form game,~$A$ a set of actions such that $A_i=A$ for all $i\in N$.  $\Gamma$ is called
	\begin{itemize}
	\item \emph{anonymous} if $p_i(a_N)=p_i(a'_N)$ for all $i\in N$ and all $a_N,a'_N\in A^N$ with $a_i=a'_i$ and $\parikh(a_{-i})=\parikh(a'_{-i})$, 
	\item \emph{symmetric} if $p_i(a_N)=p_j(a'_N)$ for all $i,j\in N$ and all $a_N,a'_N\in A^N$ with $a_i=a'_j$ and $\parikh(a_{-i})=\parikh(a'_{-j})$, 
	\item \emph{self-anonymous} if $p_i(a_N)=p_i(a'_N)$ for all $i\in N$ and all $a_N,a'_N\in A^N$ with $\parikh(a_N)=\parikh(a'_N)$, and
  \item \emph{self-symmetric}	if $p_i(a_N)=p_j(a'_N)$ for all $i,j\in N$ and all
    $a_N,a'_N\in A^N$ with $\parikh(a_N)=\parikh(a'_N)$.
	\end{itemize}
\end{definition}
When talking about anonymous games, we write $p_i(a_i,x_{-i})$ for the payoff of player~$i$ under any action profile~$a_N$ with $\parikh(a_{-i})=x_{-i}$. For self-anonymous games, $p_i(x)$ is used to denote the payoff of player~$i$ under any profile~$a_N$ with $\parikh(a_N)=x$. Unless noted otherwise, we assume that anonymous games are given explicitly, \ie as a list of payoffs for the different commutative images. 

A well-known method for simplifying strategic games is the removal of actions that are weakly dominated by some strategy of the same player, in the sense that playing the latter is never worse than playing the former and sometimes strictly better.  The removal of one or more dominated actions may render additional actions dominated, which may then iteratively be removed.
To make these notions precise, we need some notation.  Given a game $\Gamma=(N,(A_i)_{i\in N},(p_i)_{i\in N})$, call an elimination sequence of~$\Gamma$ a finite sequence $(D_1,D_2,\dots,D_k)$ of subsets of the disjoint union of the sets $A_i$, \ie $D_j\subseteq\cup_{i\in N}A_i^*$ for all $j$ with $1\leq j\leq k$, where $A_i^*=A_i\times\{i\}$.  For a set $D\subseteq\cup_{i\in N}A_i^*$, denote by $\Gamma(D)$ the induced subgame of $\Gamma$ where the actions in~$D$ have been removed, \ie $\Gamma(D)=(N,(A'_i)_{i\in N},(p_i|_{\prod_{i\in N}A'_i})_{i\in N})$ where $A'_i=\{\, a \midd (a,i)\in A_i^*\setminus D\,\}$. 
\begin{definition}[iterated dominance]\label{def:dominance}
  Let $\Gamma=(N,(A_i)_{i\in N},(p_i)_{i\in N})$ be a game.  An action $d_i\in A_i$ is said to be \emph{(weakly) dominated} by strategy $s_i\in S_i$ if for all $b\in\prod_{j\in N}A_j$, $p_i(b_{-i},d_i) \leq \sum_{a_i\in A_i}s_i(a_i)p_i(b_{-i},a_i)$ and for at least one $b\in\prod_{j\in N}A_j$, $p_i(b_{-i},d_i) < \sum_{a_i\in A_i}s_i(a_i)p_i(b_{-i},a_i)$. 
	An elimination sequence $(D_1,D_2,\dots,D_m)$ of~$\Gamma$ is called \emph{valid} if either it is the empty sequence, or if $(D_1,D_2,\dots,D_{m-1})$ is valid in~$\Gamma$ and every $d_m\in D_m$ is dominated in $\Gamma(\cup_{1\leq j\leq m-1}D_j)$. 
	An action $a\in\cup_{i\in N}A_i$ is called \emph{eliminable} if there exists a valid elimination sequence~$(D_1,D_2,\dots,D_m)$ such that~$a$ is weakly dominated in $\Gamma(\cup_{1\leq j\leq m}D_j)$.  Game~$\Gamma$ is called \emph{solvable} if it is possible to obtain a game where only one action remains for each player, \ie if there exists a valid elimination sequence~$(D_1,D_2,\dots,D_m)$ such that $\Gamma(\cup_{1\leq j\leq m}D_j)=(N,(A'_i)_{i\in N},(p'_i)_{i\in N})$ with $|A'_i|=1$ for all $i\in N$.
\end{definition}
We call iterated dominance solvability (\ids) and eliminability (\ide) the computational problems that ask for solvability of a game and eliminability of a particular action.
In contrast to iterated \emph{strict} dominance, which requires the inequality to be strict for every action profile of the other players, the result of iterated weak dominance depends on the order in which actions are removed, since the elimination of an action may render actions of another player undominated~\citep[\eg][]{Apt04a}. 

Restricted types of iterated dominance can be obtained by requiring that the dominating strategy~$s_i$ is pure, or that the elements of an elimination sequence are singletons and actions thus have to be eliminated one at a time~\citep[\eg][]{Apt04a}. 
As far as dominance by pure and mixed strategies is concerned, we will frequently exploit that the two versions coincide in games with two actions, and also in games with only two different payoffs~\citep{CoSa05b}.  All results hold for dominance by pure strategies \emph{and} for dominance by mixed strategies. 
Valid elimination sequences consisting of singletons possess a somewhat less complicated structure. We therefore in some cases restrict our attention to this specialization, and refer to the corresponding computational problems as \emph{stepwise} \ids and \ide. The results ultimately obtained for the two variants will be very similar. 
A different notion of \emph{solvability} merely requires the remaining action profiles to yield a unique payoff to each of the players~\citep[\eg][]{Moul79a}.  We note, but do not show here, that all hardness and tractability results extend to this notion as well.

\section{Complexity of Iterated Dominance}
\label{sec:initial}

Intuitively, a large number of actions neutralizes the computational advantage obtained from anonymity, by allowing for a distinction of the players by means of the \emph{actions they play}.  The search for pure Nash equilibria, for example, is tractable for anonymous games with a constant number of actions, but becomes NP-hard as soon as the number of actions grows in the number of players~\citep{BFH09a}.  In the latter case, the size of the explicit representation grows exponentially in the number of players, and one would expect natural instances of such games to be described succinctly~\citep[\cf][]{PaRo05a}.  While as a matter of fact the results of \citet{BFH09a} are established via a specific encoding of the payoff functions, namely Boolean circuits, they nevertheless provide interesting insights into the influence of restricted classes of payoff functions on the complexity of solving a game.  We give a similar result for iterated dominance in self-symmetric games, hardness for the other classes follows by inclusion. 
\begin{theorem}\label{thm:hard_growing}
  \ids and \ide are NP-hard for all four classes of anonymous games, even if the number of actions grows only logarithmically in the number of players, if only dominance by pure strategies is considered, and if there are only two different payoffs.
\end{theorem}

In the case of symmetric games, iterated dominance becomes tractable when the number of actions is bounded by a constant.
\begin{theorem}\label{thm:easy_symm}
  For symmetric and self-symmetric games with a constant number of actions, \ids and \ide can be decided in polynomial time.
\end{theorem}

In light of these two results, only one interesting class remains, namely anonymous games with a constant number of actions.  To gain a better understanding of the problem, we restrict ourselves even further to games with two actions.  It turns out that in this case iterated dominance can be reformulated in a natural way as an elimination problem on matrices.  The latter is the topic of the following section.

\section{A Matrix Elimination Problem}
\label{sec:matrix}

Let $\Gamma=([n],(\{0,1\})_{i\in N},(p_i)_{i\in N})$ be a self-anonymous game with two actions for each player, and observe that the payoffs of $\Gamma$ can be represented by a matrix~$X_{\Gamma}=(x_{i,j})_{(n+1)\times n}$ the $i$th row of which contains the payoff profile when exactly $i-1$ players play action~$1$, \ie $x_{ij}=p_j(i-1)$. It will be instructive to view iterated dominance elimination in~$\Gamma$ in terms of the corresponding operations on the matrix~$X_{\Gamma}$. For now, we restrict our attention to the case where actions are eliminated one by one, and more generally consider matrices with an arbitrary number of rows and columns. It suffices to look at matrices whose entries are natural numbers. 

Let~$X$ be an $m\times n$ matrix with entries from the natural numbers. Call a column~$c$ of~$X$ \emph{increasing} for an interval~$I$ over the rows of~$X$ if the entries in~$c$ are monotonically increasing in~$I$, with a strict increase somewhere in this interval.  Analogously, call~$c$ \emph{decreasing} for~$I$ if its entries are monotonically decreasing in~$I$, with a strict decrease somewhere in this interval.  Say that~$c$ is \emph{active} for~$I$ if it is either increasing or decreasing for this interval.  Now consider a process that starts with~$X$ and successively eliminates pairs of a row and a column.  Rows will only be eliminated from the top or bottom, such that the remaining rows always form an interval over the rows of~$X$.  A column will only be eliminated if it is active for the remaining rows.  Elimination of an increasing column is accompanied by elimination of the top row.  Analogously, a decreasing column and the bottom row are eliminated at the same time.  The process ends when no active columns remain.  

Let us define the problem more formally.  For a set~$A$, $v\in A^n$, and $a\in A$, denote by $\parikh(a,v)=|\{\,\ell\leq n \midd v_\ell=a\,\}|$ the \emph{commutative image} of~$a$ and~$v$, and write $v_{\dots k}=(c_1,c_2,\dots,c_k)$ for the prefix of~$v$ of length $k\leq n$.  Further denote $[n]=\{1,2,\dots,n\}$ and $[n]_0=\{0,1,\dots,n\}$.
\begin{definition}[matrix elimination]\label{def:matrix}
	Let $X\in\nat^{m\times n}$ be a matrix.  Call a column $k\in[n]$ of~$X$ \emph{increasing} in an interval $[i,j]\subseteq[m]$ if the sequence $x_{ik},x_{i+1,k},\dots,x_{jk}$ is monotonically increasing and $x_{ik}<x_{jk}$, \emph{decreasing} in $[i,j]\subseteq[m]$ if $x_{ik},x_{i+1,k},\dots,x_{jk}$ is monotonically decreasing and $x_{ik}>x_{jk}$, and \emph{active} if it is either increasing or decreasing.
	Then, an \emph{elimination sequence} of length~$k$ for~$X$ is a pair $(c,r)$ such that $c\in[m]^k$, $r\in\{0,1\}^k$, and for all~$i,j$ with $1\leq i<j\leq k$, $c_i\neq c_j$ and either
		$r_i=0$ and column~$c_i$ is increasing in $[\parikh(0,r_{\dots i-1})+1, m-\parikh(1,r_{\dots i-1})]$, or
		$r_i=1$ and column~$c_i$ is decreasing in $[\parikh(0,r_{\dots i-1})+1, m-\parikh(1,r_{\dots i-1})]$.
\end{definition}
 	
\begin{figure}[tb]
	\centering
  $\begin{array}{|*{4}{c|}}
    \multicolumn{1}{c}{a} & \multicolumn{1}{c}{b} & 
    \multicolumn{1}{c}{c} & \multicolumn{1}{c}{d} \\\hline
    1 & 3 & 2 & 1 \\\hline
    0 & 2 & 2 & 1 \\\hline
    0 & 2 & 3 & 0 \\\hline
    0 & 2 & 3 & 0 \\\hline
    3 & 2 & 3 & 0 \\\hline
  \end{array} 
  \hspace*{12mm}
  \begin{array}{|*{4}{c|}} 
    \multicolumn{1}{c}{a} & \multicolumn{1}{c}{b} & 
    \multicolumn{1}{c}{c} & \multicolumn{1}{c}{d} \\\hline
    1 &  & 2 & 1 \\\cline{1-1}\cline{3-4}
    0 &  & 2 & 1 \\\cline{1-1}\cline{3-4}
    0 &  & 3 & 0 \\\cline{1-1}\cline{3-4}
    0 &  & 3 & 0 \\\cline{1-1}\cline{3-4}
    \multicolumn{4}{|c|}{} \\\hline
  \end{array} 
  \hspace*{12mm}
  \begin{array}{|*{4}{c|}} 
    \multicolumn{1}{c}{a} & \multicolumn{1}{c}{b} & 
    \multicolumn{1}{c}{c} & \multicolumn{1}{c}{d} \\\hline
    \multicolumn{2}{|c|}{} & 2 & 1 \\\cline{3-4}
    \multicolumn{2}{|c|}{} & 2 & 1 \\\cline{3-4}
    \multicolumn{2}{|c|}{} & 3 & 0 \\\cline{3-4}
    \multicolumn{4}{|c|}{} \\
    \multicolumn{4}{|c|}{} \\\hline
  \end{array} 
  \hspace*{12mm}
  \begin{array}{|*{4}{c|}} 
    \multicolumn{1}{c}{a} & \multicolumn{1}{c}{b} & 
    \multicolumn{1}{c}{c} & \multicolumn{1}{c}{d} \\\hline
    \multicolumn{4}{|c|}{} \\\cline{4-4}
    \multicolumn{3}{|c|}{} & 1 \\\cline{4-4}
    \multicolumn{3}{|c|}{} & 0 \\\cline{4-4}
    \multicolumn{4}{|c|}{} \\
    \multicolumn{4}{|c|}{} \\\hline
  \end{array}$
  \caption{A matrix and a sequence of eliminations}
  \label{fig:example_matrix}
\end{figure}
Consider for example the sequence of matrices shown in \figref{fig:example_matrix}, obtained by starting with the $5\times 4$ matrix on the left and successively eliminating columns~$b$,~$a$,~$c$, and~$d$.  In this particular example, the process ends when all rows and columns of the matrix have been eliminated.  If instead we eliminated columns~$c$ and~$a$, no further eliminations would be possible.  In fact, it would be obvious after the first elimination step that we cannot obtain a sequence of length~$4$: one of the columns not eliminated so far, column~$b$, contains the same value in every row; this column cannot become active anymore, and, as a consequence, will never be eliminated.

What matters are not the actual matrix entries, but rather the difference between successive entries in a column.  A more intuitive way to look at the problem may thus be in terms of a matrix with the number of rows reduced by one, and arrows pointing downward or upward if the value increases or decreases between two adjacent entries.  A column can be deleted if it contains at least one arrow, and if all arrows in this column point in the same direction. The corresponding row to be deleted is the one at the base of the arrows.

We will be interested in two computational problems.  Matrix elimination (\me) asks whether there exists an elimination sequence that deletes the whole matrix, \ie one of length~$\min(m-1,n)$.  Eliminability of a column (\mc) is given $k\in[n]$ and asks whether there exists an elimination sequence $(c,r)$ such that for some~$i$, $c_i=k$.  Without restrictions on~$m$ and~$n$, \me and \mc turn out to be equivalent.  Indeed, both of them are equivalent to the problem of deciding whether there exists an elimination sequence eliminating certain numbers of rows from the top and bottom of the matrix.  Several other questions, like the one of an elimination sequence of a certain length, are equivalent as well.
\begin{lemma}\label{lem:me_red}
	\mc and \me are equivalent under disjunctive truth-table reductions.
\end{lemma}

When restricted to the case $m>n$, \mc is at least as hard as \me in the sense that the latter can be reduced to the former while there is no obvious reduction in the other direction.  The problem \me itself might be harder when the number of columns significantly exceeds the number of rows, because then the set of columns effectively needs to be partitioned into two sets of sizes~$m$ and $n-m$ of columns that have to be deleted and columns that can be discarded right away.

It is not hard to see that elimination of a matrix~$X$ is closely related to iterated dominance in the self-anonymous game described at the beginning of this section, where each player has two actions~$0$ and~$1$, and the payoff of player~$j$ when exactly $i-1$ players play action~$1$ is given by matrix entry $x_{ij}$.  Given actions for the other players, player~$j$ can choose between two adjacent entries of column~$j$, so one of his two actions is dominated by the other one if the column is increasing or decreasing.  Eliminating one of two actions effectively removes a player from the game, and elimination of the top or bottom row of the matrix mirrors the fact that the number of players who can still choose between both of their actions is reduced by one. Let us formally establish this relationship. 
\begin{lemma}\label{lem:iwd_red}
  Stepwise \ids and \ide in anonymous games with two actions are equivalent under disjunctive truth-table reductions to \me and \mc, respectively, restricted to instances with $m=n+1$.
\end{lemma}

We could have well allowed the simultaneous elimination of columns, and it is fairly obvious that the resulting computational problems would be equivalent to \ids and \ide. 
So why do we require columns to be eliminated one at a time? 
For one, solving \me and \mc as defined above turns out to be intricate enough to begin with, and we will ultimately not be able to characterize their complexity. 
On the other hand, the additional structure afforded by stepwise elimination will help us to gain additional insights, which we will then use to prove the main result of this paper: NP-hardness of \ids and \ide in games with three actions, both for stepwise and simultaneous eliminations. 
Finally, much of the complexity of matrix elimination already appears to be present in the stepwise version, and any result for that version can probably be extended to simultaneous eliminations as well. 

Solving \me in general turns out to be surprisingly complicated.  A natural restriction can be obtained by requiring that all columns are increasing or decreasing in~$[1,m]$.  It is not too hard to show that this makes the problem tractable irrespective of the dimensions of the matrix, and we do so in the next section as a corollary of a slightly more general result.  Unfortunately, tractability of this restricted case does not tell us a lot about the complexity of \me in general.  The latter obviously becomes almost trivial if the order of elimination for the columns is known, \ie if we are given $c\in[n]^k$ and ask whether there exists $r\in\{0,1\}^k$ such that $(c,r)$ is an elimination sequence.  This observation directly implies membership in NP.  More interestingly, deciding whether there exists $c\in[n]^k$ for a given $r\in\{0,1\}^k$ such that $(c,r)$ is an elimination sequence is also tractable.  The reason is the specific ``life cycle'' of a column.  Consider a matrix~$X$, two intervals $I,J\subseteq[m]$ over the rows of~$X$ such that $J\subseteq I$, and a column~$c\in[n]$ that is active in both~$I$ and~$J$.  Then,~$c$ must also be active for any interval~$K$ such that $J\subseteq K\subseteq I$, and~$c$ must either be increasing for all three intervals, or decreasing for all three intervals.  Thus,~$r$ determines for every $i\in[k]$ a set of possible values for $c_i$, and leaves us with a matching problem in a bipartite graph with edges in $[n]\times[k]$.  The latter can be solved in polynomial time.  Closer inspection reveals that it can in fact be decomposed into two independent matching problems on convex bipartite graphs, for which the best known upper bound is NC$^2$~\citep{Glov67a}. 

But what if nothing about $c$ and $r$ is known?  Despite the fact that we can only eliminate the top or bottom row of the matrix in each step, this still amounts to an exponential number of possible sequences.  The best upper bound for matching in convex bipartite graphs means that there currently is not much hope for constructing an algorithm that determines~$r$ nondeterministically and computes a matching on the fly.  We can nevertheless use the above reasoning to recast the problem in the more general framework of \emph{matching on paths}.  For this, we will respectively identify intervals and pairs of intervals over the rows of~$X$ with vertices and edges of a directed graph~$G$, and will then label each edge $(I,J)$ by the identifiers of the columns of~$X$ that take~$I$ to~$J$.  An elimination sequence of length~$k$ for~$X$ then corresponds to a path of length~$k$ in~$G$ which starts at the vertex corresponding to the interval~$[1,m]$, such that there exists a matching of size~$k$ between the edges on this path and the columns of~$X$.  In particular, by fixing a particular path, we obtain the bipartite matching problem described above.  A more detailed discussion of this problem is the topic of the following section.  We first study the problem itself, and return to matrix elimination toward the end of the section.

\section{Matched Paths}\label{sec:mp}

The matching problem described in the previous section generalizes the well-studied class of matching problems between two disjoint sets, or bipartite matching problems, by requiring that the elements of one of the two sets form a certain sub-structure of a combinatorial structure.  Most interesting from a computational perspective are variants where the underlying combinatorial structure can be identified in polynomial time, as it is the case for paths or for spanning trees.
\begin{definition}[matching, matched path]
  Let~$X$ be a set, $\Sigma$ an alphabet, and $\sigma:X\rightarrow 2^\Sigma$ a labeling function assigning sets of labels to elements of~$X$.  Then, a \emph{matching} of~$\sigma$ is a total function $f:X\rightarrow\Sigma$ such that for all $x,y\in X$, $f(x)\in\sigma(x)$ and $f(y)\neq f(x)$ if $y\neq x$.

  Let~$G=(V,E)$ be a directed graph, $\Sigma$ an alphabet, and $\sigma:E\rightarrow 2^{\Sigma}$ a labeling function for edges of~$G$.  Then, a \emph{matched path} of length~$k$ in~$G$ is a sequence $e_1,e_2,\dots,e_k$ such that for all~$i$ with $1\leq i<k$, there exist $u,v,w\in V$ such that $e_i=(u,v)$ and $e_{i+1}=(v,w)$, and the restriction of~$\sigma$ to $\{\,e_i\midd 1\leq i\leq k\,\}$ has a matching.
\end{definition}
We call matched path (\map) the computational problem that asks, for an explicitly given directed graph~$G$ with corresponding labeling function~$\sigma$ and an integer~$k$, whether there exists a matched path of length~$k$ in~$G$.  Variants of this problem can be obtained by asking for a matching that contains a certain set of labels, or a matched path between a particular pair of vertices.  These variants have an interesting interpretation in terms of sequencing with resources and multi-dimensional constraints on the utilization of these resources:  every resource can be used in certain states corresponding to vertices of a directed graph, and their use causes transitions between states.  The goal then is to find a sequence that uses a specific set or a certain number of resources, or one that reaches a certain state. 

In the context of this paper, we are particularly interested in instances of \map corresponding to instances of \me.  We will see later that the graphs of such instances are layered grid graphs~\citep[\eg][]{ABC+06a}, and that the labeling function satisfies a certain convexity property.  But let us look at the general problem for a bit longer.  \citet{GHR95a} consider the related \emph{labeled graph accessibility problem}, which, given a directed graph~$G$ with a single label attached to each edge, asks whether there exists a path such that the concatenation of the labels along the path is a member of a context free language~$L$ given as part of the input.  This problem is P-complete in general and LOGCFL-complete if~$G$ is acyclic.  A matching, however, corresponds to a partial permutation of the members of the alphabet, and the number of nonterminal symbols of any context-free grammar in Chomsky normal form for the permutation language over~$\Sigma$ grows super-polynomially in the size of~$\Sigma$~\citep{EKSW04a}.  It thus should not come as a surprise that the problem becomes harder when we ask for a matching.  Indeed, \map bears some resemblance to the NP-complete problem \emph{forbidden pairs} of finding a path in a directed or undirected graph if certain pairs of nodes or edges may not be used together~\citep{GMO76a}.  Instead of reducing forbidden pairs to \map, however, we show NP-hardness of a restricted version of \map using a more complicated construction, on which we will be able to build in \secref{sec:three}.  To formally state the result we need some terminology. 

Let $G=(V,E)$ be a directed graph with vertex set $V=[m]_0\times[n]_0$.  Call $(u,v)\in E$ a \emph{south edge} if for some~$i$ and~$j$, $u=(i,j)$ and $v=(i+1,j)$, and an \emph{east edge} if for some~$i$ and~$j$, $u=(i,j)$ and $v=(i,j+1)$.  Then,~$G$ is called an $m\times n$ \emph{layered grid graph} if it contains only south and east edges. 
In labeled graphs, nonexistent edges and edges that are mapped to the empty set by the labeling function are equivalent.  We therefore concentrate on \emph{complete} layered grid graphs, \ie those containing all south and all east edges. 
\begin{theorem}\label{thm:mp_hard}
	\map is NP-complete.  Hardness holds even if~$G$ is a complete layered grid graph, $|\sigma(e)|\leq 1$ for every~$e\in E$, and $|\{\,e\in E\midd\lambda\in\sigma(e)\,\}|\leq 2$ for every~$\lambda\in\Sigma$.
\end{theorem}

The proof of this theorem starts by looking at a complete $m\times n$ grid graph~$G$ for appropriate values of~$m$ and~$n$, and at a labeling function $\sigma:[m]_0\cup[n]_0\rightarrow\Sigma$.  The latter can be interpreted as a labeling function for edges of~$G$ where a label either appears on all the edges in a given row or column or on none of them.  Labels in $\Sigma$ correspond to variable occurrences in an instance of the NP-complete problem \emph{balanced one-in-three 3SAT}~\citep{Parb91a}, and~$\sigma$ is defined in such a way that a path through the graph corresponds to an assignment of truth values to variable occurrences.  The overall structure of the graph consist of two parts. In the first part, consistency of the overall assignment is ensured by placing labels corresponding to different occurrences of the same variable on the same path.  In the second part, the same labels are used again to verify that all clauses are satisfied by the assignment. 
To obtain \thmref{thm:mp_hard} and get a better understanding of the minimal requirements for hardness, the graph is then modified further. An important property of the labeling function in this context seems to be that the same label can appear at least twice in different parts of the graph. 

The labeling function~$\sigma$ can also more generally be interpreted as belonging to a more general graph where transitions can take place from any vertex to any other vertex to the south and east of it, as long as the distance in columns between the two vertices is at most the number of unused labels that appear on the row associated with the former vertex, and the same condition holds for the distance in rows and the number of labels on the column.  Intuitively, this type of transition occurs when several dominated actions of a game are eliminated simultaneously. It will play an important role in the proof of~\thmref{thm:hard_three}. 

\begin{figure}[tb]
  \centering
	\input{example_graph}
	\caption{Labeled graph for the matrix elimination instance of \figref{fig:example_matrix}. A matched path and its matching are shown in bold.}
	\label{fig:example_graph}
\end{figure}
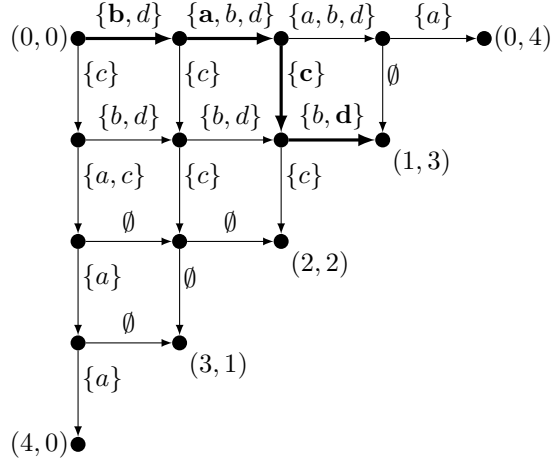
Let us now return to matrix elimination.  In light of \thmref{thm:mp_hard}, an efficient algorithm for \me would have to exploit additional structure of \map instances induced by instances of \me.  This structure is indeed quite restricted in that edges carrying a particular label~$\lambda$ satisfy a ``directed'' convexity condition: if~$\lambda$ appears on two edges $e=(u,v)$ and $e'=(u',v')$, then~$\lambda$ must appear on \emph{all} south edges or on all east edges that lie on a path from~$u$ to~$v'$, but not both. In particular, if there is such a path, it cannot be that one of~$e$ and~$e'$ is a south edge and the other is an east edge.  This fact is illustrated in \figref{fig:example_graph}, which shows the labeled graph for the \me instance of \figref{fig:example_matrix}, as well as a matched path corresponding to an elimination sequence of maximum length.

\begin{definition}[directed convexity]
  Let $G=(V,E)$ be a complete layered grid graph.  A labeling function $\sigma:E\rightarrow 2^{\Sigma}$ for~$G$ is called \emph{directed convex} if for every label $\lambda\in\Sigma$ and for every set of three edges $e_1=(u_1,v_1)$, $e_2=(u_2,v_2)$, $e_3=(u_3,v_3)$, such that $u_2$ is reachable from~$u_1$,~$u_3$ is reachable from~$u_2$, and $\lambda\in\sigma(e_1)\cap\sigma(e_3)$, it holds that~$e_1$ and $e_3$ have the same direction and $\lambda\in\sigma(e_2)$ if and only if $e_2$ has the same direction as well. 
\end{definition}

It is not too hard to see that instances corresponding to \me have a directed convex labeling function. 
\begin{lemma}\label{lem:mp_red}
	\me is polynomial time many-one reducible to \map restricted to layered grid graphs and directed convex labeling functions.  
\end{lemma}

Directed convexity of the labeling function means that we cannot show NP-hardness of \me by a construction similar to the one used in the proof of \thmref{thm:mp_hard}.  On the other hand, it is not quite clear how the additional structure provided by directed convexity can be exploited to obtain a polynomial-time algorithm for \me.  The case $m\leq n$ will probably add additional complications. We therefore leave the complexity of \me as an open problem, albeit quite an elegant one.  

Here we consider a more special case of \map, which provides additional insights.  In the corresponding instances of \me, all columns are active at the beginning of the matrix elimination process, or all columns are active in the interval of length one at the end of the elimination process.
\begin{definition}[backward and forward closure]
  Let $G=(V,E)$ be a complete layered grid graph.  Let~$s$ be the unique vertex of $G$ with in-degree zero,~$t$ the unique vertex with outdegree zero.  Then, a labeling function $\sigma:E\rightarrow 2^{\Sigma}$ for~$G$ is called \emph{backward closed} if $\{ \lambda\in\sigma(s,v) \midd (s,v)\in E \}=\Sigma$.  Similarly,~$\sigma$ is called \emph{forward closed} if $\{ \lambda\in\sigma(s,v) \midd (v,t)\in E \}=\Sigma$.
\end{definition}

It may not have gone unnoticed that these properties are closely related to closure properties found respectively in matroids and antimatroids.  Together with directed convexity, each of the closure properties further implies that each label appears only on east edges or only on south edges.  This allows us to consider two distinct matching problems, one for east and one for south edges, and obtain a tractability result. 
\begin{theorem} \label{thm:active_easy}
	Let $G=(V,E)$ be a complete layered grid graph,~$\sigma$ a labeling function for~$G$ that is directed convex and either backward or forward closed.  Then, \map for $G$ and $\sigma$ can be solved in nondeterministic logarithmic space.
\end{theorem}

A generalization of both backward and forward closure can be obtained by considering labeling functions that are \emph{connected} in the sense that the edges carrying a particular label, together with all edges in the respective other direction, form a weakly connected graph.  This property introduces a dependence between the matching problems for the two directions, and a very interesting question is whether \thmref{thm:active_easy} can be generalized to this setting.


\section{Self-Anonymous Games with a Constant Number of Actions}\label{sec:three}

It is natural to ask whether iterated dominance for games with more than two actions can still be interpreted in terms of eliminations in a matrix or matrix-like structure.  Consider a self-anonymous game with~$\numactions{}$ actions.  As before, the payoff of a particular player~$i$ only depends on the number of players, including the player itself, that play each of the different actions.  They can thus be written down as entries in a discrete simplex of dimension~$k-1$.  The elimination of the $\ell$th action by some player can then be interpreted as a cut along the~$\ell$th $0$-face of the simplex of every player.

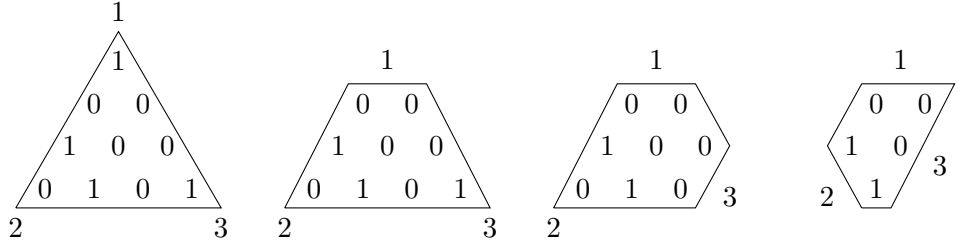
\begin{figure}[tb]
	\centering
	\input{example_triangle}
	\caption{Payoffs of a particular player in a self-anonymous game with~$\numplayers{}=3$ and~$\numactions{}=3$. Initially all actions are pairwise undominated. If one of the other players eliminates action~$1$, action~$3$ weakly dominates action~$1$. Action~$1$ then becomes undominated if some player deletes action~$3$, and dominated by action~$2$ if one more player deletes action~$3$, and some player deletes action~$2$.}
	\label{fig:example_triangle}
\end{figure}

The left hand side of \figref{fig:example_triangle} shows the payoffs of a particular player in a self-anonymous game with~$\numplayers{}=3$ and~$\numactions{}=3$.
Compared to matrix elimination as introduced in \defref{def:matrix} and illustrated in \figref{fig:example_matrix}, we notice an interesting shift, which curiously has nothing to do with the added possibility of dominance by mixed strategies.  Rather, a particular action $a\in A$ may now be eliminated by either one of several other actions in~$A\setminus\{a\}$, and the situations where~$a$ can be eliminated no longer form a convex set. 

This already indicates that it might be possible to construct a layered grid graph with corresponding labeling function for which the existence of a matched path is NP-hard to decide, and which is induced by a self-anonymous game with three actions for each player.  To obtain our main result we however have to overcome one additional obstacle: when dropping the assumption that actions are eliminated one at a time, the equivalence between elimination sequences and labeled paths in a layered grid graph breaks down. We therefore start from the construction used in the proof of \thmref{thm:mp_hard}, and use additional vertices and labels to make it work for the more general type of transitions corresponding to the simultaneous elimination of actions. 
\begin{theorem}\label{thm:hard_three}
	\ids and \ide are NP-complete.  Hardness holds even for self-anonymous games with three actions and two different payoffs, and also applies to stepwise \ids and \ide. 
\end{theorem}

\acknowledgements
This material is based upon work supported by the Deutsche Forschungsgemeinschaft under grants BR~2312/3-1 and BR~2312/3-2.  We thank Hermann Gruber, Paul Harrenstein, Tim Roughgarden, Inbal Talgam, and Michael Tautschnig for valuable discussions, and apologize to Edith Hemaspaandra for spoiling the sunset at White Sands.



\end{document}

%% file: example_graph.tex
\begin{tikzpicture}[scale=.9] 
	\small 
	\tikzstyle{every circle node}=[fill,inner sep=2pt,outer sep=0pt]
  \path (0,0) node[circle](v04){} node[left]{$(0,0)$} ++(1.5,0) node[circle](v03){} ++(1.5,0) node[circle](v02){} ++(1.5,0) node[circle](v01){} +(1.5,0)  node[circle](v00){} node[right]{$(0,4)$} ++(0,-1.5) node[circle](v11){} node[below right]{$(1,3)$} ++(-1.5,0) node[circle](v12){} ++(-1.5,0) node[circle](v13){} ++(-1.5,0) node[circle](v14){} ++(0,-1.5) node[circle](v24){} ++(1.5,0) node[circle](v23){} +(1.5,0) node[circle](v22){} node[below right]{$(2,2)$} ++(0,-1.5) node[circle](v33){} node[below right]{$(3,1)$} ++(-1.5,0) node[circle](v34){} ++(0,-1.5) node[circle](v44){} node[left]{$(4,0)$};
  \draw[-latex,very thick] (v04) -- node[above]{$\{\mathbf{b},d\}$} (v03);
  \draw[-latex,very thick] (v03) -- node[above]{$\{\mathbf{a},b,d\}$} (v02);
  \draw[-latex] (v02) -- node[above]{$\{a,b,d\}$} (v01);
  \draw[-latex] (v01) -- node[above]{$\{a\}$} (v00);
	\draw[-latex] (v04) -- node[right=-2pt,pos=.33]{$\{c\}$} (v14);
  \draw[-latex] (v03) -- node[right=-2pt,pos=.33]{$\{c\}$} (v13);
  \draw[-latex,very thick] (v02) -- node[right=-2pt,pos=.33]{$\{\mathbf{c}\}$} (v12);
  \draw[-latex] (v01) -- node[right=-2pt,pos=.33]{$\emptyset$} (v11);
	\draw[-latex] (v14) -- node[above]{$\{b,d\}$} (v13);
  \draw[-latex] (v13) -- node[above]{$\{b,d\}$} (v12);
  \draw[-latex,very thick] (v12) -- node[above]{$\{b,\mathbf{d}\}$} (v11);
  \draw[-latex] (v14) -- node[right=-2pt,pos=.33]{$\{a,c\}$} (v24);
  \draw[-latex] (v13) -- node[right=-2pt,pos=.33]{$\{c\}$} (v23);
  \draw[-latex] (v12) -- node[right=-2pt,pos=.33]{$\{c\}$} (v22) ;
  \draw[-latex] (v24) -- node[above]{$\emptyset$} (v23);
  \draw[-latex] (v23) -- node[above]{$\emptyset$} (v22);
  
  \draw[-latex] (v24) -- node[right=-2pt,pos=.33]{$\{a\}$} (v34);
  \draw[-latex] (v23) -- node[right=-2pt,pos=.33]{$\emptyset$} (v33);
  \draw[-latex] (v34) -- node[above]{$\emptyset$} (v33);
  \draw[-latex] (v34) -- node[right=-2pt,pos=.33]{$\{a\}$} (v44);
\end{tikzpicture}

%% file: example_triangle.tex
\begin{tikzpicture}[scale=0.65] 
  \path (0,0) node{$0$} +(-.6,-.4) coordinate(a) node[below]{$2$} ++(1,0) node{$1$} ++(1,0) node{$0$} ++(1,0) node{$1$} +(.6,-.4) coordinate(b) node[below]{$3$} ++(-.5,.87) node{$0$} ++(-1,0) node{$0$} ++(-1,0) node{$1$} ++(.5,.87) node{$0$} ++(1,0) node{$0$} ++(-.5,.87) node{$1$} +(0,.6) coordinate(c) node[above]{$1$};
	\draw (a) -- (b) -- (c) -- (a);
	\path (5.5,0) node{$0$} +(-.6,-.4) coordinate(d) node[below]{$2$} ++(1,0) node{$1$} ++(1,0) node{$0$} ++(1,0) node{$1$} +(.6,-.4) coordinate(e) node[below]{$3$} ++(-.5,.87) node{$0$} ++(-1,0) node{$0$} ++(-1,0) node{$1$} ++(.5,.87) node{$0$} +(-.3,.4) coordinate(f) ++(1,0) node{$0$} +(.3,.4) coordinate(g) +(-.5,.5) node[above]{$1$};
	\draw (d) -- (e) -- (g) -- (f) -- (d);
	\path (11,0) node{$0$} +(-.6,-.4) coordinate(h) node[below]{$2$} ++(1,0) node{$1$} ++(1,0) node{$0$} +(.3,-.4) coordinate(i) ++(.5,.87) node{$0$} +(.5,0) coordinate(j) ++(-1,0) node{$0$} ++(-1,0) node{$1$} ++(.5,.87) node{$0$} +(-.3,.4) coordinate(k) ++(1,0) node{$0$} +(.3,.4) coordinate(l)  +(-.5,.5) node[above]{$1$};
	\draw (h) -- (i) -- node[below right]{$3$} (j) -- (l) -- (k) -- (h);
	\path (16,0) ++(1,0) node{$1$} +(-.3,-.4) coordinate(m) +(.3,-.4) coordinate(n) ++(.5,.87) node{$0$} ++(-1,0) node{$1$} +(-.5,0) coordinate(o) ++(.5,.87) node{$0$} +(-.3,.4) coordinate(p) ++(1,0) node{$0$} +(.6,.4) coordinate(q) +(-.5,.5) node[above]{$1$};
	\draw (m) -- (n) -- node[below right]{$3$} (q) -- (p) -- (o) -- node[below left]{$2$} (m);
\end{tikzpicture}

%% file: brandt.bbl
\begin{thebibliography}{19}
\providecommand{\natexlab}[1]{#1}
\providecommand{\url}[1]{\texttt{#1}}
\expandafter\ifx\csname urlstyle\endcsname\relax
  \providecommand{\doi}[1]{doi: #1}\else
  \providecommand{\doi}{doi: \begingroup \urlstyle{rm}\Url}\fi

\bibitem[Allender et~al.(2006)Allender, {Mix Barrington}, Chakraborty, Datta,
  and Roy]{ABC+06a}
E.~Allender, D.~A. {Mix Barrington}, T.~Chakraborty, S.~Datta, and S.~Roy.
\newblock Grid graph reachability problems.
\newblock In \emph{Proceedings of the 21st Annual IEEE Conference on
  Computational Complexity (CCC)}, pages 299--313, 2006.

\bibitem[Apt(2004)]{Apt04a}
K.~R. Apt.
\newblock Uniform proofs of order independence for various strategy elimination
  procedures.
\newblock \emph{Contributions to Theoretical Economics}, 4\penalty0 (1), 2004.

\bibitem[Brandenburger et~al.(2008)Brandenburger, Friedenberg, and
  Keisler]{BFK08a}
A.~Brandenburger, A.~Friedenberg, and H.~J. Keisler.
\newblock Admissibility in games.
\newblock \emph{Econometrica}, 76\penalty0 (2):\penalty0 307--352, 2008.

\bibitem[Brandt et~al.(2009{\natexlab{a}})Brandt, Brill, Fischer, and
  Harrenstein]{BBFH09b}
F.~Brandt, M.~Brill, F.~Fischer, and P.~Harrenstein.
\newblock On the complexity of iterated weak dominance in constant-sum games.
\newblock In M.~Mavronicolas and V.~G. Papadopoulou, editors, \emph{Proceedings
  of the 2nd International Symposium on Algorithmic Game Theory (SAGT)}, volume
  5814 of \emph{Lecture Notes in Computer Science (LNCS)}, pages 287--298.
  Springer-Verlag, 2009{\natexlab{a}}.

\bibitem[Brandt et~al.(2009{\natexlab{b}})Brandt, Fischer, and Holzer]{BFH09a}
F.~Brandt, F.~Fischer, and M.~Holzer.
\newblock Symmetries and the complexity of pure {N}ash equilibrium.
\newblock \emph{Journal of Computer and System Sciences}, 75\penalty0
  (3):\penalty0 163--177, 2009{\natexlab{b}}.

\bibitem[Conitzer and Sandholm(2005)]{CoSa05b}
V.~Conitzer and T.~Sandholm.
\newblock Complexity of (iterated) dominance.
\newblock In \emph{Proceedings of the 6th ACM Conference on Electronic Commerce
  (ACM-EC)}, pages 88--97. ACM Press, 2005.

\bibitem[Daskalakis and Papadimitriou(2008)]{DaPa08a}
C.~Daskalakis and C.~H. Papadimitriou.
\newblock Discretized multinomial distributions and {N}ash equilibria in
  anonymous games.
\newblock In \emph{Proceedings of the 49th Symposium on Foundations of Computer
  Science (FOCS)}. IEEE Computer Society Press, 2008.

\bibitem[Ellul et~al.(2004)Ellul, Krawetz, Shallit, and Wang]{EKSW04a}
K.~Ellul, B.~Krawetz, J.~Shallit, and M.-W. Wang.
\newblock Regular expressions: {N}ew results and open problems.
\newblock \emph{Journal of Automata, Languages and Combinatorics}, 9\penalty0
  (2--3):\penalty0 233--256, 2004.

\bibitem[Gabow et~al.(1976)Gabow, Maheshwari, and Osterweil]{GMO76a}
H.~N. Gabow, S.~N. Maheshwari, and L.~Osterweil.
\newblock On two problems in the generation of program test paths.
\newblock \emph{IEEE Transactions on Software Engineering}, 2\penalty0
  (3):\penalty0 227--231, 1976.

\bibitem[Gilboa et~al.(1993)Gilboa, Kalai, and Zemel]{GKZ93a}
I.~Gilboa, E.~Kalai, and E.~Zemel.
\newblock The complexity of eliminating dominated strategies.
\newblock \emph{Mathematics of Operations Research}, 18\penalty0 (3):\penalty0
  553--565, 1993.

\bibitem[Glover(1967)]{Glov67a}
F.~Glover.
\newblock Maximum matching in convex bipartite graphs.
\newblock \emph{Naval Research Logistics Quarterly}, 14:\penalty0 313--316,
  1967.

\bibitem[Greenlaw et~al.(1995)Greenlaw, Hoover, and Ruzzo]{GHR95a}
R.~Greenlaw, H.~J. Hoover, and W.~L. Ruzzo.
\newblock \emph{Limits to Parallel Computation}.
\newblock Oxford University Press, 1995.

\bibitem[Knuth et~al.(1988)Knuth, Papadimitriou, and Tsitsiklis]{KPT88a}
D.~E. Knuth, C.~H. Papadimitriou, and J.~N. Tsitsiklis.
\newblock A note on strategy elimination in bimatrix games.
\newblock \emph{Operations Research Letters}, 7:\penalty0 103--107, 1988.

\bibitem[Moulin(1979)]{Moul79a}
H.~Moulin.
\newblock Dominance solvable voting schemes.
\newblock \emph{Econometrica}, 47:\penalty0 1337--1351, 1979.

\bibitem[Myerson(1991)]{Myer91a}
R.~B. Myerson.
\newblock \emph{Game Theory: Analysis of Conflict}.
\newblock Harvard University Press, 1991.

\bibitem[Papadimitriou and Roughgarden(2005)]{PaRo05a}
C.~H. Papadimitriou and T.~Roughgarden.
\newblock Computing equilibria in multi-player games.
\newblock In \emph{Proceedings of the 16th Annual ACM-SIAM Symposium on
  Discrete Algorithms (SODA)}, pages 82--91. SIAM, 2005.

\bibitem[Parberry(1991)]{Parb91a}
I.~Parberry.
\newblock On the computational complexity of optimal sorting network
  verification.
\newblock In \emph{Proceedings of the Conference on Parallel Architectures and
  Languages Europe (PARLE)}, volume 505 of \emph{Lecture Notes in Computer
  Science (LNCS)}, pages 252--269. Springer-Verlag, 1991.

\bibitem[Parikh(1966)]{Pari66a}
R.~Parikh.
\newblock On context-free languages.
\newblock \emph{Journal of the ACM}, 13\penalty0 (4):\penalty0 570--581, 1966.

\bibitem[Samuelson(1992)]{Samu92a}
L.~Samuelson.
\newblock Dominated strategies and common knowledge.
\newblock \emph{Games and Economic Behavior}, 4:\penalty0 284--313, 1992.
\end{thebibliography}
